\documentclass{article}
\usepackage{amsmath}
\usepackage{amssymb}
\usepackage{graphicx}
\usepackage{xcolor}
\usepackage{siunitx}
\usepackage{subcaption}
\usepackage{textcomp}
\usepackage{enumitem}
\usepackage{booktabs}
\usepackage{float}
\usepackage{authblk}
\usepackage{hyperref}
\usepackage{cleveref}
\usepackage{bbm}
\usepackage{algorithm}
\usepackage{multirow}
\usepackage{amsmath}
\usepackage{amssymb}
\usepackage{graphicx}
\usepackage{booktabs}
\usepackage{multirow}
\usepackage{caption}
\usepackage[margin=1in]{geometry}
\usepackage[round]{natbib}
\usepackage[toc,page]{appendix}
\makeatletter
\renewcommand\hyper@natlinkbreak[2]{#1}
\makeatother

\graphicspath{ {./figs/} }

\newcommand{\mupop}{\mu_{\mathrm{pop}}}
\newcommand{\sigpop}{\sigma_{\mathrm{pop}}}
\newcommand{\sigmeas}{\sigma_{\mathrm{meas}}}

\newcommand{\sigtot}{\sigma_{\mathrm{total}}}
\newcommand{\thetapop}{\theta_{\mathrm{pop}}}
\newcommand{\thetameas}{\theta_{\mathrm{meas}}}
\newcommand{\fpop}{f_{\mathrm{pop}}}
\newcommand{\gmeas}{g_{\mathrm{meas}}}

\graphicspath{ {../figs/} } 

\makeatother

\def\bSig\mathbf{\Sigma}

\def\spacingset#1{\renewcommand{\baselinestretch}%
{#1}\small\normalsize} \spacingset{1}

\begin{document}

  \title{\bf When to Repeat a Biomarker Test? A Framework for Decisions Under
Measurement Error }

\author[1]{Supun Manathunga}
\author[2]{Mart P. Janssen}
\author[3]{Yu Luo}
\author[4]{W. Alton Russell}
\author[2]{Mart Pothast}


\renewcommand\Affilfont{\small}
\affil[1]{{Experimental Medicine}, {McGill University}, {{Montreal}, {Canada}}}
\affil[2]{{Transfusion Technology Assessment}, {Sanquin Research}, {{Amsterdam}, {Netherlands}}}
\affil[3]{{Department of Mathematics}, {King's College}, {{London}, {United Kingdom}}}
\affil[4]{{Epidemiology, Biostatistics and Occupational Health}, {McGill University}, {{Montreal}, {Canada}}}
  \date{\today}
  \maketitle


\abstract
{Repeating a noisy measurement of a continuous biomarker can reduce the risk of misclassification. In many clinical settings, measurements are only repeated if they fall within a certain range. Measurement variability can be untangled from population variability using data with multiple measurements per individual, but only if all individuals are equally likely to be retested. In this paper, we show how conditional retesting biases naive estimates of measurement error. We demonstrate how Bayesian hierarchical modelling can untangle measurement error from population biomarker variability despite conditional retesting and yield unbiased probability estimates that an individual’s biomarker falls within a clinically relevant range based on one or more noisy measurements. Applying our method to blood donor records, we estimate the contribution of measurement error to overall variability in pre-donation fingerstick hemoglobin measurements. We demonstrate how donor eligibility algorithms that employ conditional retesting alter the trade-offs between two types of classification error: ‘false bleeds’ in which low-hemoglobin donors are deemed eligible and ‘false deferrals’ in which donors with sufficient hemoglobin are turned away.}

\noindent%

\vfill

\newpage
\spacingset{1.5}

\section{Introduction}
\label{s:intro}

Biomarker levels, such as blood glucose, C-reactive protein, and hemoglobin, play a prominent role in modern medicine. Diagnosis and treatment decisions often involve a biomarker and decision threshold to classify an individual as positive for a condition (e.g., diagnose diabetes based on hemoglobin A1C) or as indicated for an intervention (e.g., transfuse red cells based on hemoglobin). In the presence of significant measurement uncertainty, a single determination may misclassify a non-negligible fraction of the individuals whose values lie near the threshold.
To avoid this, clinicians may repeat a borderline or unexpected measurement. For instance, it is common to repeat clinical chemistry measurements in a `critical range', though this practice is contested \citep{Niu2013,Soleimani2021,Sun2018}. We refer to this practice of selective retesting based on an initial measurement as `conditional retesting'. Policies using conditional retesting are typically applied to reduce a specific misclassification risk, but may increase the risk of another (e.g., reduce the false positive rate while increasing the false negative rate).

Conditional retesting gives rise to routinely collected data in which multiple measurements are only available for select individuals. In this paper, we show how such data can be used to quantify the contribution of measurement error, as opposed to population variability (i.e. variability between individuals). Once estimated, our model can be used to estimate how conditional retesting policies alter risk of misclassification due to measurement error. 

\subsection{Related prior work and limitations}

In the absence of a gold standard comparator, the variance of a measurement procedure is conventionally estimated from repeated measurements on the same individuals, by halving the variance of their
differences \citep{BlandAltman1996}. This calculation, and the wider literature on measurement-error models \citep{Carroll2006,Wang2024}, treats the repeated observations as unconditionally observed. When observation of a repeat measurement depends on the value of the initial result, the retested pairs are a selected subset and will give a biased estimate of the measurement variance. 

Problems of similar nature have been studied in other settings. Estimation after a group sequential test in clinical trials carries a bias conditional on the stopping stage, and conditional maximum likelihood estimators have been derived for it \citep{Liu2004,Whitehead1986}. These treatments mainly concern a general shift of the location parameter or a treatment effect, but we are interested in quantifying the error of the measurement.

Bayesian hierarchical models have been used to separate measurement-level noise from individual-level
heterogeneity. \citet{gustafson2003measurement} provides treatment for measurement
error and misclassification, showing how a hierarchical structure can carry
uncertainty in both the exposure assessment and the outcome process.
\citet{greenland2005multiple} embeds such a model in a multiple-bias framework,
adjusting simultaneously for exposure misclassification, selection bias and
confounding. \citet{luo2018estimating} uses a hierarchical Bayesian model to
estimate prevalence while propagating variability, spatial dependence and
estimation uncertainty into the reported figure. However, these methodologies do not directly deal with conditionally repeated measurements. 

Deciding whether to repeat a measurement, and how to handle the resulting measurements, can be informed by the probability that an individual's latent (true) value lies above the threshold given the measurements recorded for that individual.
Simply estimating the variance alone is not sufficient to determine that probability, which depends on
the shapes of the population and error densities and not only on their variances. We explore the ability to determine the full probability structure using data on blood donor eligibility screening, where a donor is accepted only above a hemoglobin threshold and a low reading is often repeated.

\subsection{Hemoglobin screening in blood donors}

Most countries screen blood donors against a minimum hemoglobin (Hb) threshold before donation, as recommended by the  World Health Organization, \citep{WorldHealthOrganization2012}, generally using a fingerstick capillary sample with a point-of-care device \citep{Zalpuri2020}. 
Low Hb is the most common reason for deferring a donor on site \citep{VeldHuizen2010}. This screening can protect donors from donation-induced iron deficiency, which is associated with low Hb at subsequent donations \citep{Prinsze2021}.
However, deferral also costs a donation and consumes resources, causes inconvenience to a donor who has traveled to donate and may reduce the likelihood of return \citep{Bruhin2020}.

Fingerstick Hb measurements vary considerably and drop-to-drop variation exceeds that
of venous samples \citep{Bond2015,Hackl2024,Killilea2022}. Many deferrals are the result of an erroneously low reading, which we term a `false deferral' \citep{Janssen2022}. Other donors may be inaccurately classified as having sufficient Hb and are allowed to donate, which we call `false bleeds' \citep{Bell2021}. Because it is assumed that false deferrals are more common than false bleeds, many blood donation centers repeat a below-threshold measurement \citep{Zalpuri2020}. However, there is no clear evidence to which retesting rule is best. Recording the maximum of the available measurements introduces bias, as shown by \citet{Chung2019}, and distorts the recorded distribution \citep{Pothast2025}. Because a repeat is recorded only after a low first result, these data have the selective structure as described above.

\subsection{Contribution and outline}
We develop a flexible Bayesian hierarchical model to jointly estimate the population and measurement-error distributions from conditionally repeated point-of-care fingerstick hemoglobin measurements. The proposed framework yields posterior probabilities that an individual’s true hemoglobin level exceeds a pre-specified threshold, conditional on one or more observed measurements. We use these probabilities to evaluate retesting policies by quantifying the trade-off between unnecessarily deferring donors whose true hemoglobin levels are adequate and accepting donors whose levels are below the threshold.

\Cref{sec:problem} contains the variance decomposition for unconditionally repeated measurements and conditional retesting under the assumption of normality. In \cref{sec:methods} we describe the Bayesian model, establish that the retesting mechanism is ignorable when inferring this model and define the predictive criterion used to
compare candidate distributions. In \cref{sec:simulation_study} we use simulations to test how well that criterion
recovers the generating probability distributions and what is the impact of the retest probability.
In \cref{sec:case_study} the framework is applied to fingerstick hemoglobin measurements from United States blood donors. \Cref{sec:discussion} considers what the results mean for repeat testing in this setting.

\section{Problem formulation}
\label{sec:problem}

\subsection{Notation and variance decomposition}

We assume that the total variation of a biomarker measurement stems from two sources, the variation of the true biomarker level between individuals and the variation introduced by the measurement procedure.

We define the true biomarker level for individual $i$ as $T_i$, drawn from the population distribution with mean $\mupop$ and noise $\epsilon_{\mathrm{pop},i}$, so that
\[
T_i=\mupop+\epsilon_{\mathrm{pop},i}.
\]
A measurement $x_i$ with noise $\epsilon_{\mathrm{meas},i}$ is then
\[
    x_i = T_i + \epsilon_{\mathrm{meas},i}
        = \mupop + \epsilon_{\mathrm{pop},i} + \epsilon_{\mathrm{meas},i}.
\]
We assume both noise terms have mean zero and define
$\sigpop^2=\mathrm{Var}(\epsilon_{\mathrm{pop},i})$ and
$\sigmeas^2=\mathrm{Var}(\epsilon_{\mathrm{meas},i})$. Assuming the two are
independent, the probability distribution of measurements is the convolution
\begin{equation}    
f_X(x) = \left(\fpop * \gmeas\right)(x)
           = \int \fpop(t)\,\gmeas(x-t)\,dt,
\label{eq:convolution}
\end{equation}
where $\fpop$ is the density of biomarker levels in the population and $\gmeas$
the density of the measurement error for a single measurement.

For an individual with repeated measurements, measured $n_i$ times, the $j$-th measurement is 
\begin{equation} 
x_{i,j}=T_i+\epsilon_{i,j}
\label{eq:meas_true}
\end{equation}
 where for brevity we from now on use $\epsilon_{i,j}=\epsilon_{\mathrm{meas},i,j}$. 

The usual way to determine the measurement error from two measurements per individual is by determining the difference $\Delta_i=x_{i,1}-x_{i,2}=\epsilon_{i,1}-\epsilon_{i,2}$. If the
errors are independent and identically distributed and independent of $T_i$,
then for an unconditionally observed sample
\begin{equation}
\mathrm{Var}(\Delta_i) = 2\sigmeas^2,\ \sigmeas^2 = \mathrm{Var}(\Delta_i)/2,
\label{eq:non-cond-sigma}
\end{equation}
so the measurement variance can be estimated from repeated measurements
\citep{BlandAltman1996}. The variance of a single measurement
decomposes as
\begin{align}
    \sigtot^2 = \mathrm{Var}(x_{i,j}) = \sigpop^2 + \sigmeas^2.
    \label{eq:total_variance}
\end{align}
\Cref{eq:total_variance} requires only
that the errors are uncorrelated with $T_i$ and with each other, so it holds for
any $\fpop$ and $\gmeas$ with finite variance.

\subsection{Conditional retesting under normality}
\label{subsec:conditional-normal}

Under \textit{conditional retesting}, the decision to take a second measurement
depends on the first. A second fingerstick hemoglobin measurement, for example,
may be recorded only when a first measurement falls below the
eligibility threshold. Let $c$ denote a cut-off threshold, such that a repeat is
recorded only when $x_{i,1}<c$, and let $\mathcal{R}$ index the retested
individuals with $M=|\mathcal{R}|$.

In this subsection, we assume normal distributions such that
$T_i\overset{\mathrm{iid}}{\sim}\mathcal{N}(\mupop,\sigpop^2)$ and
$\epsilon_{i,j}\overset{\mathrm{iid}}{\sim}\mathcal{N}(0,\sigmeas^2)$, with the
true values independent of the errors. In that case the measurement pair $(x_{i,1},x_{i,2})$, when recorded for every individual, is bivariate normal with variance $\sigtot^2$ and correlation $\rho=\sigpop^2/\sigtot^2$. With \cref{eq:total_variance}, estimating $\sigpop^2$ and $\sigmeas^2$ is equivalent to estimating $\sigtot^2$ and $\rho$. It is possible to show the resulting bias when the second measurement is repeated conditionally, shown in \cref{s:bias}. 

The first measurements of the pairs are observed unconditionally, so their sample mean and
standard deviation can be used to estimate $\mupop$ and $\sigtot$. Holding these
fixed and standardizing the measurements by them, one can write the conditional likelihood of
the retested pairs and estimate $\rho$ from solving a cubic equation. In \cref{s:rho} this equation is derived and shown that the relevant root in $(-1,1)$ exists. 

This analytical solution for the measurement variability, while informative under normality assumptions, does not allow estimating misclassification risks under conditional retesting policies. This is addressed in the next section using Bayesian hierarchical models, which allow for non-normal distributions for $\fpop$ and $\gmeas$. In addition to estimating measurement variability, the hierarchical models can generate posterior predictions on a person's true biomarker based on one or more noisy measurements, which are easily converted into  misclassification risks for a given retesting policy.

\section{Bayesian hierarchical model for conditionally repeated measurements}
\label{sec:methods}

Whether an individual's true biomarker level is above a threshold is a statement about the latent value $T_i$ given observed measurement. A hierarchical model can estimate these probabilities from data resulting from a conditional retesting policy. Inference for the parameters of $\fpop$ and $\gmeas$
requires integrating over individuals' latent (true) biomarker levels $T_i$, and posterior sampling carries out
that integration while returning draws of the latent values themselves.

\subsection{Hierarchical measurement-error model}

As before, we assume an underlying population distribution
$\fpop$ of the latent true biomarker levels for individuals $i=1,\ldots,N$. For
individual $i$, let $n_i$ denote the number of measurements recorded in order
as $x_{i,1},\ldots,x_{i,n_i}$. Write
$\mathbf{x}_{\mathrm{obs}}=\{x_{i,j}:i=1,\ldots,N,\ j=1,\ldots,n_i\}$ for the
collection of recorded measurements. Each measurement satisfies \cref{eq:meas_true}, where the errors follow the distribution $\gmeas$. We assume that, conditional on the model parameters, the latent biomarker levels are
independent across individuals, measurement errors are independent across individuals and between repeated measurements on the same individual, and decisions to repeat a measurement are independent of the latent biomarker level given already-observed measurements. The population and measurement-error distributions are governed by parameters $\thetapop$ and $\thetameas$, respectively, which have prior distributions.

The data generation process can then be summarized in the following hierarchy:
\begin{align*}
\text{Priors: }& \theta_{\mathrm{pop}} \sim p(\theta_{\mathrm{pop}}),\theta_{\mathrm{meas}} \sim p(\theta_{\mathrm{meas}}),\\
\text{Population model: }& T_i \mid \theta_{\mathrm{pop}} \sim f_{\mathrm{pop}}(\,\cdot \mid \theta_{\mathrm{pop}}),\\
\text{Measurement model: }& x_{i,j}-T_i \mid \theta_{\mathrm{meas}}
\sim g_{\mathrm{meas}}(\,\cdot \mid \theta_{\mathrm{meas}}).
\end{align*}
Integrating out $T_i$ for an individual measured once returns the convolution
of \cref{eq:convolution}.

The posterior for $\mathbf{T}=\{T_i\}_{i=1}^n$, $\thetapop$ and $\thetameas$,
with prior distributions $p(\thetapop), p(\thetameas)$, given the recorded
measurements is proportional to
\begin{equation}
p(\mathbf{T}, \theta_{\mathrm{pop}}, \theta_{\mathrm{meas}}
\mid \mathbf{x}_{\mathrm{obs}})
\propto
p(\theta_{\mathrm{pop}})
\, p(\theta_{\mathrm{meas}})
\prod_{i=1}^{N}
f_{\mathrm{pop}}(T_i \mid \theta_{\mathrm{pop}}) \prod_{i=1}^{N}\prod_{j=1}^{n_i} g_{\mathrm{meas}}(x_{i,j}-T_i \mid \theta_{\mathrm{meas}}).
\label{eq:posterior}
\end{equation}
One can integrate over latent values $T_i$ to get the marginal distribution of
$\thetapop$ and $\thetameas$
\begin{equation}
\begin{split}
p(\theta_{\mathrm{pop}}, \theta_{\mathrm{meas}}
\mid \mathbf{x}_{\mathrm{obs}})
\propto
p(\theta_{\mathrm{pop}})
\, p(\theta_{\mathrm{meas}})
\prod_{i=1}^{N}
\int
f_{\mathrm{pop}}(T_i \mid \theta_{\mathrm{pop}}) 
\prod_{j=1}^{n_i}
g_{\mathrm{meas}}(x_{i,j}-T_i \mid \theta_{\mathrm{meas}})
\, dT_i.
\end{split}
\label{eq:marginal_posterior}
\end{equation}

\subsection{Conditionally repeated measurements}
\label{subsec:conditional_repeats}

Here we assume that a first measurement is always recorded and at
most one repeat is taken (though the framework can be expanded to consider multiple repeats per individuals). Let $R_i\in\{0,1\}$ indicate whether the second
measurement on individual $i$ is recorded, so that $n_i=1+R_i$, and write
$\mathbf{R}=(R_1,\ldots,R_N)$. The measurement model defines $x_{i,2}$ for
every individual, but it is recorded only when $R_i=1$. We assume that the probability of a retest
\[
\Pr\!\left(R_i=1\mid
x_{i,1},x_{i,2},T_i,\thetapop,\thetameas,\psi\right)
=h_2(x_{i,1};\psi),
\]
where $h_2(\cdot;\psi)$ gives the retesting probability and $\psi$ denotes its
parameters (scalar or vector), which are distinct from $\thetapop$ and
$\thetameas$ and a priori independent of them.
Retesting decisions are thus only dependent on the first measurement, which is always observed, and so the mechanism is missing at random \citep{little2019statistical}.

In a commonly used threshold-based protocol, a repeat is indicated only when the
first measurement falls below a known cutoff $c$. A simple special case is
\[
h_2(x_{i,1};\psi) = \psi\,\mathbbm{1} (x_{i,1} < c),
\]
where $0<\psi\leq1$ is the probability that an indicated repeat is carried
out, and $\psi=1$ recovers the deterministic protocol. In this rule, an
indicated repeat may be omitted, but gives probability zero to a repeat at or
above $c$, and the retested index set of \cref{subsec:conditional-normal} is
$\mathcal{R}=\{i:R_i=1\}$.

\subsection{Flexible distributional specifications}
\label{subsec:specifications}

The framework places no restriction on $\fpop$ and $\gmeas$ beyond tractable
densities and a measurement error centered at zero. To assess departures from normality in the population and measurement error distributions, we consider four models with different specification for $T_i$ and $\epsilon_{i,j}$.
\texttt{Model a} assumes both population and measurement error distribution to be normal, 
\[
T_i \sim \mathcal N(\mupop,\sigpop^2), \qquad \epsilon_{i,j} \sim \mathcal N(0,s^2),
\] 
the specification under which the estimators of \cref{subsec:conditional-normal} are valid. \texttt{Model b} keeps the normal population and takes Student-$t$ measurement errors, 
\[
T_i \sim \mathcal N(\mupop,\sigpop^2), \qquad \epsilon_{i,j} \sim t_\nu(0,s), 
\]
where $s$ is the scale parameter and $\nu$ the degrees of freedom. \texttt{Model c} also keeps the normal population distribution with measurement errors following a two-component normal mixture,
\[
T_i \sim \mathcal N(\mupop,\sigpop^2), \qquad \epsilon_{i,j} \sim (1-w)\,\mathcal N(0,s_1^2) + w\,\mathcal N(0,s_2^2),
\]
where a proportion $w$ of the measurements carries the larger scale $s_2$. \texttt{Model d} pairs the Student-$t$ measurement error with a skew-normal population distribution,
\[
T_i \sim \mathrm{SN}(\xi,\omega,\alpha), \qquad \epsilon_{i,j} \sim t_\nu(0,s),
\]
where $\xi$, $\omega$, and $\alpha$ denote the location, scale, and shape parameters of the population distribution, respectively. For $\alpha=0$, the population distribution is normal, so that \texttt{Model d} reduces to \texttt{Model b}.

Outside the normal case the scale parameters are not standard deviations.
Writing $\sigmeas^2=\mathrm{Var}(\epsilon_{i,j})$ as in \cref{sec:problem},
\[
\sigmeas^2 =
\begin{cases}
s^2 & (\texttt{a}),\\
s^2\nu/(\nu-2) & (\texttt{b},\texttt{d}),\\
(1-w)s_1^2 + w s_2^2 & (\texttt{c}),
\end{cases}
\]
the Student-$t$ variance requiring $\nu>2$, and for the skew-normal population,
with $\delta=\alpha/\sqrt{1+\alpha^2}$, $\mupop = \xi + \omega\delta\sqrt{2/\pi}, 
\sigpop^2 = \omega^2\left(1-2\delta^2/\pi\right)$. The decomposition of
\cref{eq:total_variance} therefore holds under all four, and their variance
components can be compared directly.

\subsection{Predictive model comparison}
\label{subsec:model_selection}

We compare models \texttt{a}--\texttt{d} by their out-of-sample predictive
performance, fitting each to a training set and scoring it on a held-out set.
The true value $T_i$ of a held-out individual is not in the training
posterior, so for each posterior draw from the training fit it is integrated
out as in \cref{eq:marginal_posterior} \citep{Merkle2019}. Averaging the
result over draws gives the predictive density of the individual's recorded
measurements. The integral over $T_i$ is computed by Gauss--Hermite quadrature
when it has no closed form. A model's expected log predictive density (elpd) is
estimated by
\[
\mathrm{elpd}=\sum_{i}\log p\left(x_{i,1},\ldots,x_{i,n_i}\mid \mathbf{x}_{\mathrm{train}}\right),
\]
where the sum is taken over the held-out individuals and $\mathbf{x}_{\mathrm{train}}$
denotes the training data. The retest indicators are also part of the held-out data, but their contribution is
the same under all four specifications because all four use the same retesting
model (\cref{s:ignorability}).

Every model is scored on the same set of individuals, so the difference in elpd
between two models is a sum of individual-level differences. Its paired
standard error is the standard deviation of these differences multiplied by the
square root of the number of held-out individuals \citep{Vehtari2017}.

When no separate validation set is available, we use cross-validation, where individuals are split into folds, with all of an individual's measurements in
the same fold. Each fold is scored by the model fitted to the remaining
folds. Every individual is scored once, and all specifications use the
same folds. The simulation study (\cref{sec:simulation_study}) scores an
independent validation set and the case study (\cref{sec:case_study}) uses
five-fold cross-validation.

\section{Simulation study}
\label{sec:simulation_study}

\subsection{Design}

We use simulation to answer two questions about our framework: (1) does the model selection procedure from  \cref{subsec:model_selection} consistently select the correct data generating distributions, and (2) are estimates of measurement variability robust to lower retesting probabilities.

Loosely based on hemoglobin levels among blood donors, all data-generating processes share $\mupop =15$~g/dL,
$\sigpop =1$~g/dL and $\sigmeas =0.6$~g/dL, with a retesting threshold $c = 14.4$~g/dL. The distinguishing
features are a normal error in \texttt{Model a}, a Student-$t$ error with degrees of freedom $\nu=3$ in 
\texttt{Model b}, a normal mixture error with $w=0.1$, $s_1=0.283$ and $s_2=1.697$~g/dL in
\texttt{Model c}, and population skewness $\alpha=6$ with a Student-$t$ error with $\nu=5$ in
\texttt{Model d}. Exact parameters, priors and computational settings are given
in \cref{tab:sim-settings}.

For evaluating model selection, 50 independent datasets were generated under each
specification, each with 1,000 individuals for training and a further 1,000 for
validation. Each candidate model specification was fitted to each training dataset and scored
on the corresponding validation dataset. The retesting rule was
deterministic, $R_i=\mathbbm{1}(x_{i,1}<14.4)$, so $\psi=1$.

For evaluating lower retesting probabilities, the rule was
$h_2(x_{i,1};\psi)=\psi\,\mathbbm{1}(x_{i,1}<14.4)$ with $\psi$ taking the
value in $\{0.1, 0.5, 0.8, 1\}$. Twenty-five independent datasets were
generated for each combination of $\psi$ and generating specification, and the
correctly specified model was fitted to each.

\subsection{Recovery of the generating specification}
\Cref{tab:selection-frequency} reports how often the generating specification achieved the largest elpd across the 50 datasets, and \cref{tab:selection-matrix} gives the full counts of which specification was selected. Models \texttt{c} and \texttt{d} were recovered almost always, model \texttt{a} in 78 per cent of datasets and model \texttt{b} in 46 per cent.

Model \texttt{a} was never selected unless it generated the data. Most errors under model \texttt{b} were selections of model \texttt{d}, which reduces to model \texttt{b} at $\alpha=0$. In each of these datasets the central 90\% credible interval for $\alpha$ contained zero, so the selected model \texttt{d} was close to model \texttt{b}. When model \texttt{d} generated the data, with $\alpha=6$, every interval excluded zero. The second column of \cref{tab:selection-frequency} gives the fraction of datasets in which the elpd of the generating specification was within one paired standard error of the best, as in the one-standard-error rule \citep{Breiman1984}. Used as a selection procedure, choosing the simplest specification within one paired standard error of the best, the rule recovered the generating specification in 184 of the 200 datasets (\cref{tab:selection-one-se}).

\begin{table}[t]
\centering
\caption{Fraction of 50 simulated datasets in which the generating
specification had the largest elpd, or an elpd within one paired standard
error of the largest.}
\label{tab:selection-frequency}
\begin{tabular}{lcc}
\toprule
Generating model & Largest elpd (SE) & Within one SE \\
\midrule
\texttt{a} (normal, normal)      & 0.78 (0.06) & 1.00 \\
\texttt{b} (normal, $t$)         & 0.46 (0.07) & 0.88 \\
\texttt{c} (normal, mixture)     & 0.98 (0.02) & 1.00 \\
\texttt{d} (skew-normal, $t$)    & 1.00 (0.00) & 1.00 \\
\bottomrule
\end{tabular}
\end{table}

\subsection{Sensitivity to the retest probability}
Across all four retest probabilities, the measurement standard deviation was recovered with no detectable bias (posterior mean within two Monte Carlo standard errors of the truth in 15
of the 16 combinations; \cref{tab:psi-precision})). Its Monte Carlo standard
error increased by a factor of $1.5$ to $3.2$ when decreasing $\psi$ from 1 to 0.1. Fewer
repeats therefore reduce precision but do not introduce bias, as expected based on the ignorability argument
of \cref{subsec:conditional_repeats} \cref{s:ignorability}.

The degrees of freedom $\nu$ were overestimated at every retest probability, so
the fitted measurement errors had lighter tails than the truth. The bias grew as
$\psi$ fell for model \texttt{b}, suggesting that with fewer repeats the posterior for $\nu$ relies more on the prior.
Coverage stayed near the nominal level (\cref{tab:psi-precision}).

No fit had a divergent transition, but fits with fewer repeats mixed less well,
with E-BFMI below 0.3 \citep{Betancourt2017} in 74 per cent of fits at $\psi=0.1$ and 12 per cent at
$\psi=1$. At $\psi=1$, refitting model \texttt{a} with its latent values
integrated out exactly changed the posterior means by at most 0.03 posterior
standard deviations and the held-out elpd by 0.056, against differences of 1 to
15 between candidate specifications.

\section{Case study: blood donor hemoglobin screening}
\label{sec:case_study}

We apply the framework to the screening of blood donors with fingerstick Hb measurements, where a second measurement is taken and recorded when the first falls below the eligibility cutoff. The case study selects the population and measurement-error distributions by held-out predictive performance, estimates the share of observed Hb variation due to measurement error, and examines what the fitted model implies for individual eligibility decisions and retesting policies.

\subsection{Data and analysis cohort}
The data were provided by Vitalant, a large blood collector in the United
States, and covered donation visits from January 2017 through October 2022.
The source data comprised 9,099,136 visits from 2,582,402 donors, of which
6,527,996 visits had a first fingerstick Hb measurement.

\begin{figure}[t]
    \centering
    \includegraphics[width=\linewidth]{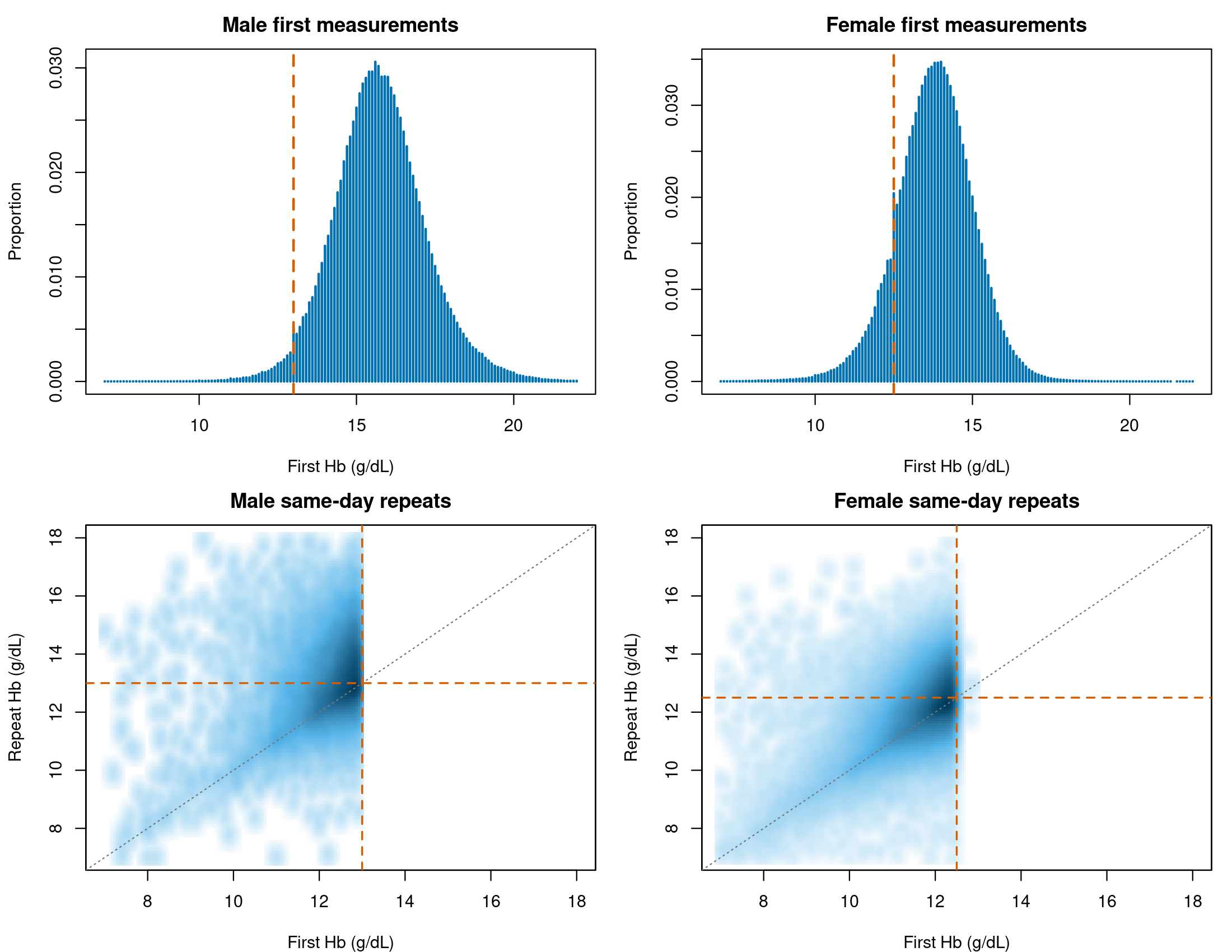}
    \caption{Fingerstick Hb measurements in the unique-donor first-visit
    cohort. Top: distributions of first measurements. Bottom: joint
    distributions of same-day first and repeat measurements among retested
    donors. Orange dashed lines mark the sex-specific eligibility cutoffs;
    grey dotted lines mark equality of the measurements.}
    \label{fig:data_hists}
\end{figure}

To avoid dependence across visits and changes in Hb over time, we retained only the
first visit for each donor, which included one or two recorded Hb measurements. The resulting cohort contained 1,863,158 donors,
The eligibility cutoff was $13\,\mathrm{g/dL}$ for males and $12.5\,\mathrm{g/dL}$ for females, and $c$ denotes the cutoff for the donor's sex from here on.
Among males, 18,172 (2.1\%) first measurements were strictly below $c$, of
which 16,556 (91.1\%) had a same-day repeat. Among females, 123,381 (12.2\%)
first measurements were below $c$, of which 111,331 (90.2\%) had a same-day
repeat. A further 12 repeats were recorded at or above $c$, giving 111,343
observed female repeats in total. Thus, repeats were observed almost
exclusively after a below-cutoff first result, as in the conditional-retesting
design of \cref{subsec:conditional_repeats}. Not every indicated repeat
was recorded; possible reasons include policy deviation, documentation failure, or that the donor declined undergoing a second measurement. The full selection flow appears in
\cref{fig:selection-flow}; \cref{fig:data_hists} shows the distribution of Hb measurements in the analysis
cohort.

\subsection{Implementation and predictive comparison}
\label{subsec:case-model-selection}
We fitted each of the four specifications from \cref{subsec:specifications} as a single model for both sexes, with separate population and measurement-error parameters for males and females and the priors in \cref{tab:priors-real}. The mixture weight
in model \texttt{c} was shared across sexes.

Predictive comparison used a random sample of 20,000
donors, with 10,000 per sex and 1,264 same-day repeats. We formed five folds at
the donor level, stratifying by sex, repeat status, and whether $x_{i,1}<c$;
every model used the same folds. Models \texttt{b} and \texttt{d} used
100-node quadrature for fitting and 200-node quadrature for held-out scoring.
Per-fold and total elpds are given in \cref{tab:cv-folds}.

Model \texttt{d}, with a skew-normal population and Student-$t$ measurement error, had the largest elpd in every fold. Its total elpd exceeded that of model \texttt{b} by 58.36 (paired SE 11.83), model \texttt{c} by 101.95 (SE 30.69) and model \texttt{a} by 365.31 (SE 108.10), and the approximate 95\% intervals for all three differences excluded zero (\cref{tab:cv_diff}). Model \texttt{b} ranked second. The standardized skewness of model \texttt{d} was small (0.23 for males and $-0.24$ for females). We therefore retained model \texttt{b}, whose
normal population distribution is more parsimonious, for the primary analysis.


For primary inference, we fitted model \texttt{b} to an independent sample of
100,000 donors, balanced by sex and containing 6,381 same-day repeats. Because
many donors shared identical observed-data patterns, we evaluated each distinct
combination of sex, repeat status, \(x_{i,1}\), and, when observed, \(x_{i,2}\)
only once. Let \(\mathbf{y}_p\) denote the \(p\)th of \(P\) distinct patterns
and \(N_p\) its frequency. We therefore replaced $\sum_{i=1}^{N}\log L_i(\theta)$ with $\sum_{p=1}^{P}N_p\log L(\mathbf{y}_p\mid\theta)$.
This reduced the final 100,000-donor sample to 1,795 likelihood patterns. Four
chains each used 1,000 warm-up and 1,000 retained MCMC iterations, 100-node
quadrature, target average proposal acceptance probability of $0.99$, and maximum tree depth 12. The final
fit passed all diagnostic criteria (\cref{fig:trace}). The maximum
$\widehat R$ was 1.005, the minimum bulk and tail effective sample sizes were
1,541 and 2,012, respectively, and there were no divergences or
maximum-tree-depth transitions.

\subsection{Population and measurement variability}
The estimated latent population distributions differed by sex
(\cref{tab:posterior_summary}). The posterior mean of $\mupop$ was
$15.736\,\mathrm{g/dL}$ for males and $13.790\,\mathrm{g/dL}$ for females;
the corresponding population SDs were $1.286$ and $1.083\,\mathrm{g/dL}$.

The Student-$t$ scales were similar for males and females (0.359 and
$0.357\,\mathrm{g/dL}$), as were the measurement-error IQRs (0.561 and
$0.543\,\mathrm{g/dL}$). The lower degrees of freedom for males, however,
indicated heavier measurement-error tails. Measurement error accounted
for posterior mean fractions of 25.9\% and 23.1\% of total variance among males
and females, respectively.

\begin{table}[t]
\centering
\caption{Final model \texttt{b} posterior means with central 90\% credible intervals.
The measurement-error SD is $s\sqrt{\nu/(\nu-2)}$, the IQR is
$2s\,t_{\nu,0.75}$, and the variance fraction is the measurement share of
$\sigtot^2$.}
\label{tab:posterior_summary}
\begin{tabular}{lrrrrrr}
\toprule
& \multicolumn{3}{c}{Male} & \multicolumn{3}{c}{Female} \\
\cmidrule(lr){2-4}\cmidrule(lr){5-7}
Parameter & Mean & 5\% & 95\% & Mean & 5\% & 95\% \\
\midrule
$\mupop$ (g/dL) & 15.74 & 15.73 & 15.75 & 13.79 & 13.78 & 13.80 \\
$\sigpop$ (g/dL) & 1.29 & 1.28 & 1.30 & 1.08 & 1.07 & 1.09 \\
$s$ (g/dL) & 0.36 & 0.34 & 0.38 & 0.36 & 0.35 & 0.37 \\
$\nu$ & 2.59 & 2.44 & 2.74 & 3.14 & 2.99 & 3.31 \\
Measurement-error SD (g/dL) & 0.76 & 0.72 & 0.81 & 0.59 & 0.58 & 0.61 \\
Measurement-error IQR (g/dL) & 0.56 & 0.53 & 0.59 & 0.54 & 0.53 & 0.56 \\
Measurement variance fraction (\%) & 26 & 24 & 29 & 23 & 22 & 24 \\
\bottomrule
\end{tabular}
\end{table}

\subsection{Individual eligibility probabilities}
\begin{figure}[t]
    \centering
    \includegraphics[width=1\linewidth]{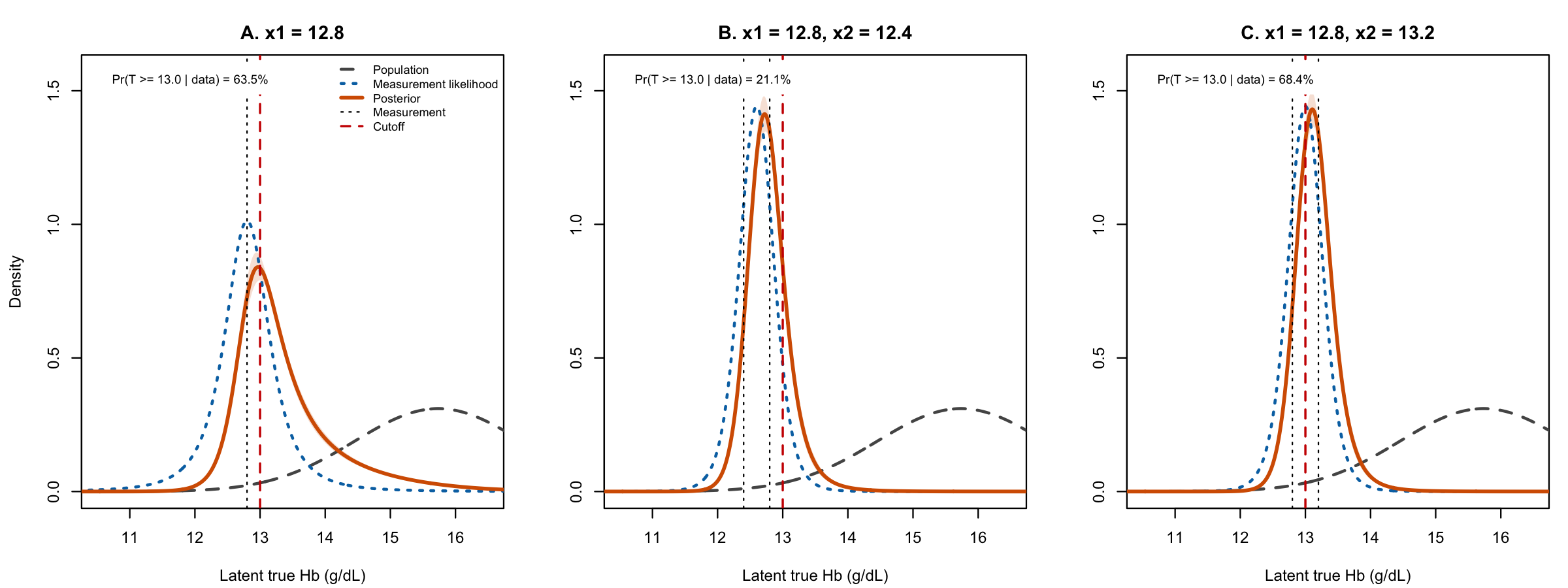}
    \caption{Fitted population density, normalized measurement likelihood,
    and posterior latent-Hb density for three male examples with a
    $13\,\mathrm{g/dL}$ cutoff. Panel A uses one measurement of
    $12.8\,\mathrm{g/dL}$; panel B adds $12.4\,\mathrm{g/dL}$; panel C adds
    $13.2\,\mathrm{g/dL}$. Shading of the posterior curve gives the pointwise central 90\%
    interval for the density.}
    \label{fig:app-examples}
\end{figure}

For each global posterior draw, we integrated the conditional distribution of
$T_i$ to obtain the probability that a donor's same-visit Hb exceeded the
cutoff. For example, a male donor with $x_{i,1}=12.8\,\mathrm{g/dL}$ has a
63.5\% posterior probability of $T_i\geq13\,\mathrm{g/dL}$ (central 90\%
credible interval 61.2\%--65.8\%). A second measurement of
$12.4\,\mathrm{g/dL}$ reduces this probability to 21.1\% (18.7\%--23.5\%),
whereas a second measurement of $13.2\,\mathrm{g/dL}$ increases it to 68.4\%
(67.5\%--69.3\%; \cref{fig:app-examples}).

\subsection{Screening and retesting policies}
We compared following screening policies by their posterior rates of false deferral
(FD), deferring a donor with $T_i\geq c$, and false bleed (FB), accepting a
donor with $T_i<c$, both as percentages of all donors, and by the expected
number of measurements per donor.
\begin{itemize}
\item \textit{Single measurement}, accepting a donor when $x_{i,1}\geq c$.
\item \textit{Shifted single threshold}, accepting a donor when $x_{i,1}$ was at
or above a threshold varied from $c-2$ to $c+2$ g/dL.
\item \textit{Mean of two} and \textit{maximum of two}, taking two measurements
from every donor and accepting when their mean, or their maximum, was at or
above $c$.
\item \textit{Conditional maximum}, obtaining $x_{i,2}$ only when $x_{i,1}<c$
and accepting when either measurement was at or above $c$. This corresponds to
the retesting in the case-study data.
\item \textit{Adaptive posterior}, with an uncertainty band $(l,u)$ and
$q_j=\Pr(T_i\geq c\mid x_{i,1},\ldots,x_{i,j})$ computed at the posterior means
of the model parameters. A donor was deferred if $q_1\leq l$, accepted if $q_1\geq u$, and
otherwise retested and accepted if $q_2\geq0.5$.
\end{itemize}

\begin{figure}[t]
    \centering
    \includegraphics[width=\linewidth]{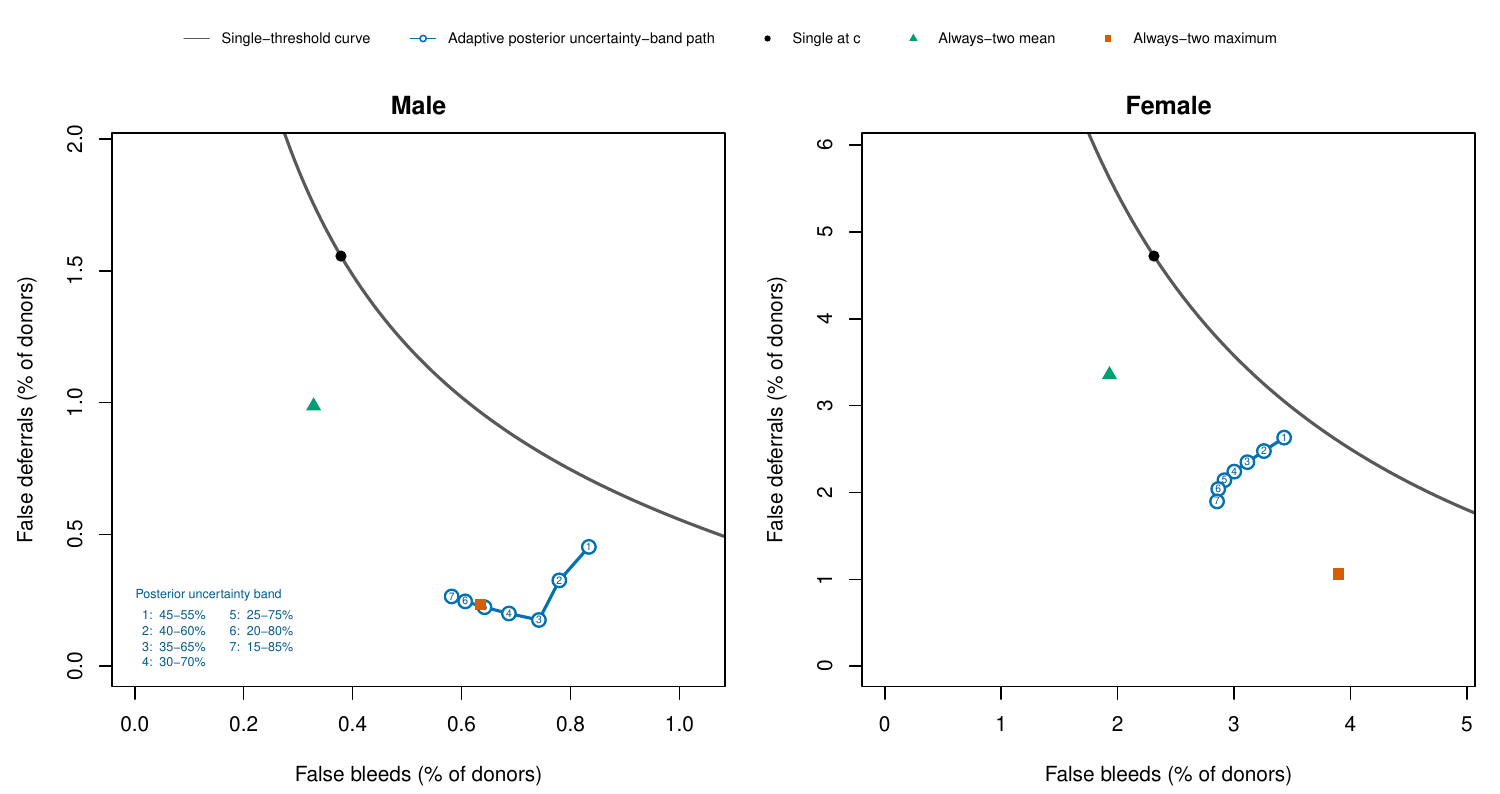}
    \caption{Posterior mean false-bleed and false-deferral rates under Model
    \texttt{b}, shown over the policy-relevant region. The black curve varies a
    single-test threshold from $c-2$ to $c+2$ g/dL; the black point uses $c$.
    The green triangle and orange square use the mean and maximum of two
    measurements, respectively; conditional maximum coincides with the orange
    square. The numbered blue points vary the posterior uncertainty band that
    triggers a repeat, as defined by the key. Lower and further left indicate
    fewer errors of both types; measurement use is reported in
    \cref{tab:policy-comparison}.}
    \label{fig:policy-tradeoff}
\end{figure}

Operating characteristics were evaluated over all 4,000 posterior draws.
\Cref{fig:policy-tradeoff} shows FD against FB for every policy, and
\cref{tab:policy-comparison} gives the rates with credible intervals and the
expected number of measurements.

Relative to a single measurement, the conditional maximum reduced FD from
1.56\% to 0.23\% among males and from 4.72\% to 1.06\% among females, but
increased FB from 0.38\% to 0.63\% and from 2.31\% to 3.90\%, respectively
(\cref{tab:policy-comparison}). Under a single measurement, 54.7\% of deferred
males and 33.5\% of deferred females had eligible latent Hb, and under the
conditional maximum these shares fell to 18.4\% and 12.0\%. The conditional
maximum made the same decisions as the maximum of two while taking a second measurement for only 2.85\% of male and
14.08\% of female donors. The mean of the two was the only policy that
reduced both errors relative to a single measurement, but it required two
measurements from every donor. The adaptive policies gave intermediate
trade-offs between errors and measurement use. The 25--75\% band, for example,
required 1.026 measurements per male and 1.100 per female donor.

The shifted single threshold did not reach the operating points of the two-measurement policies. At the same FD as the mean of two, the matched single threshold produced FBs of 0.620\% versus 0.328\% for
males and 3.169\% versus 1.930\% for females. At the same FB, it produced FDs
of 1.752\% versus 0.987\% and 5.620\% versus 3.357\%, respectively. The same
pattern held for the maximum of two, and all paired central 90\% posterior
intervals for the excess error under a matched single threshold excluded
zero. Moving one threshold therefore changes the balance between the two errors,
whereas a second measurement adds information and reaches operating points
below the single-threshold curve.

These comparisons do not identify a clinically optimal policy without
relative costs for FD, FB, and another fingerstick.


\begin{table}[t]
\centering
\caption{Posterior mean operating characteristics, 90\% credible intervals in
parentheses. FD and FB are percentages of all donors and
$E(N_{\mathrm{meas}})$ the expected measurements per donor. The 25--75\% and
15--85\% labels are the first-measurement posterior bands that trigger
retesting.}
\label{tab:policy-comparison}
\begin{tabular}{llccc}
\toprule
Sex & Policy & FD (\%) & FB (\%) & $E(N_{\mathrm{meas}})$ \\
\midrule
\multirow{6}{*}{Male}
 & Single at $c$ & 1.56 (1.48--1.63) & 0.38 (0.37--0.39) & 1 \\
 & Always two, mean & 0.99 (0.94--1.03) & 0.33 (0.32--0.34) & 2 \\
 & Always two, maximum & 0.23 (0.22--0.25) & 0.63 (0.61--0.66) & 2 \\
 & Conditional maximum & 0.23 (0.22--0.25) & 0.63 (0.61--0.66) & 1.028 (1.028--1.029) \\
 & Posterior, 25--75\% & 0.22 (0.21--0.24) & 0.64 (0.62--0.66) & 1.026 (1.025--1.027) \\
 & Posterior, 15--85\% & 0.26 (0.25--0.28) & 0.58 (0.56--0.60) & 1.033 (1.032--1.034) \\
\addlinespace
\multirow{6}{*}{Female}
 & Single at $c$ & 4.72 (4.61--4.83) & 2.31 (2.27--2.36) & 1\\
 & Always two, mean & 3.36 (3.28--3.43) & 1.93 (1.90--1.97) & 2 \\
 & Always two, maximum & 1.06 (1.03--1.09) & 3.90 (3.83--3.98) & 2 \\
 & Conditional maximum & 1.06 (1.03--1.09) & 3.90 (3.83--3.98) & 1.141 (1.139--1.143) \\
 & Posterior, 25--75\% & 2.14 (2.08--2.20) & 2.92 (2.87--2.97) & 1.100 (1.099--1.101) \\
 & Posterior, 15--85\% & 1.90 (1.85--1.95) & 2.86 (2.80--2.91) & 1.166 (1.165--1.168) \\
\bottomrule
\end{tabular}
\end{table}

\section{Discussion}
\label{sec:discussion}

Conditionally repeating a continuous biomarker test introduces a form of
sequential testing bias. This paper shows how data arising from such a process
can still be used to separate the contribution of the measurement procedure
from the variation between individuals, and to quantify the risk of
misclassification under policies using one or more measurements.

The usual difference-based estimator of measurement variability is biased when repeats
are conditional. Under the assumption of normality the repeated measurements follow a multivariate normal distribution which can be used to infer the variation from the measurement. However, this breaks down for more complicated distributional specifications.

Therefore, we specify the population and measurement-error distributions explicitly in a Bayesian hierarchical framework. We use simulations to show that it is possible to infer the correct specification using the expected log predictive density even if the probability of retesting is varied. Applying our method to hemoglobin measurements prior to blood donations, we find that a heavy-tailed measurement-error distribution fits the data best and that about a quarter of the total variance is from the measurement procedure. 

Routine repeat testing of Hb measurements below the donation threshold is intended
to reduce false deferrals caused by measurement error. By modelling explicitly the population and measurement variation we show that it
also increases the number of donors accepted whose latent true hemoglobin is below
the threshold, trading one error against the other. We also studied other possible strategies and their effect on false deferrals and false bleeds. The mean of two measurements reduces both misclassification errors, but requires two measurements for every donor. Moving the
single-measurement threshold cannot reproduce what a second measurement
achieves, because a threshold shift moves along a fixed error trade-off while
an extra measurement adds information. Using an ``adaptive posterior'' policy, which uses the posterior probability and an uncertainty band to determine wether to retest, the number of false deferrals and false bleeds can be reduced in an explainable way, while preventing unnecessary repeat measurements. 

Several limitations may be addressable through extensions of this framework. First, we ignored within-person fluctuation in biomarker levels. Extending the hierarchical framework to include a within-person variance component is conceptually straightforward, and would be appropriate for  settings where
measurements occur on different days. Second, we assume the measurement error has mean zero and is independent of the latent level. Hence, no systematic over- or underestimation is incorporated and this ignores any variability that varies with the biomarker level. Third, our case study conditioned only on sex and on measurements
taken at one point in time. A model accounting for other covariates influencing hemoglobin, such as donation history, previous biomarker values, and weight, may reduce posterior uncertainty, enabling adaptive policies that further reduce misclassification error.

With the estimated measurement error and population distribution it is possible to make a
probabilistic assessment of donor eligibility. One or two fingerstick
measurements become a posterior probability about a donor's true latent hemoglobin
relative to the threshold (\cref{fig:app-examples}). This makes explicit when a
second measurement would meaningfully change the decision and when it would add
little. For a donor whose first measurement is far from the threshold, a repeat
changes little. For a donor near the threshold, it can reverse the decision. The posterior rule directs the repeat by how uncertain the decision is rather than by whether the first measurement crossed the threshold, and it reached the operating points on fewer repeats than either fixed policy
that uses a second measurement.

In summary, this paper provides a framework for separating population variation
from measurement error under a selective retesting protocol and for converting
the resulting estimates into misclassification probabilities, supporting decision making.
Applied to blood donor hemoglobin screening in the United States, it yields
estimates of measurement and population variability, an explicit account of what
repeat testing costs and gains and a basis for evaluating candidate policies
that does not depend on assuming the measurement error is normal.








\section*{Acknowledgments}
The authors thank the blood donors whose data enabled this study. The authors
also thank collaborators Ralph Vassallo and Marjorie Bravo from Vitalant
Medical Affairs and Brian Custer and Zhanna Kaidarova from Vitalant Research
Institute for sharing data and providing feedback on our study. The authors used Claude for language editing and assistance with writing code. The authors reviewed, edited, and approved all output and take full responsibility for the content.

\section*{Study Funding}

This research was funded in part by the Natural Sciences and Engineering
Research Council of Canada (NSERC) [funding reference number RGPIN-2023-04160,
PI W. Alton Russell].

\section*{Conflict of interest}

The authors declare no potential conflict of interests.

\section*{Data and Code Availability}
The analysis code is available at
\url{https://github.com/ssm123ssm/Hb-variability---code.git}. The repository
includes scripts that simulate data and apply the methods described in the
manuscript. Blood donor data were analyzed under a data sharing agreement and
ethics approval that do not permit sharing of individual-level data.

\bibliographystyle{plainnat}\bibliography{library}

\begin{appendices}
\crefalias{section}{appendix}

\section{Bias of the naive difference estimator}
\label{s:bias}

Let $x_{i,j}=T_i+\epsilon_{i,j}$ for $N$ individuals and $j=1,2$, with
$T_i\sim\mathcal N(\mupop,\sigpop^2)$ and
$\epsilon_{i,j}\sim\mathcal N(0,\sigmeas^2)$, all independent.
$T_i$ is the true value, $\epsilon_{i,j}$ is measurement error,
$\mupop$ and $\sigpop^2$ are the population mean and variance, and
$\sigmeas^2>0$ is the error variance. Set
$\sigtot^2=\sigpop^2+\sigmeas^2$.

For a fixed, finite threshold $c$, only individuals satisfying
$S_i=\{x_{i,1}<c\}$ are eligible for retesting. Each donor is retested independently
with probability $\psi\in(0,1]$, independently of true levels and errors within
this group. Let $\mathcal R$ be the retested set and $M=|\mathcal R|$. Define
$c^{\star}=(c-\mupop)/\sigtot$, $\kappa=\sigmeas^2/\sigtot^2$ and
$\lambda(u)=\phi(u)/\Phi(u)$, where $\phi$ and $\Phi$ are the standard normal
density and cumulative distribution function.

The distribution of retested differences $\Delta_i=x_{i,1}-x_{i,2}=\epsilon_{i,1}-\epsilon_{i,2}$ is conditional on $S_i$ and since $S_i$ depends only on
$(T_i,\epsilon_{i,1})$, the second error remains independent with its original
distribution. Therefore,
\[
\mathrm{E}(\Delta_i\mid S_i)=\mathrm{E}(\epsilon_{i,1}\mid S_i),
\qquad
\mathrm{Var}(\Delta_i\mid S_i)=\mathrm{Var}(\epsilon_{i,1}\mid S_i)+\sigmeas^2.
\]
$\mathrm{Cov}(\epsilon_{i,1},x_{i,1})=\sigmeas^2$ and the first error can
be written $\epsilon_{i,1}=\kappa(x_{i,1}-\mupop)+r_i$, where the residual
$r_i$ is normal with mean zero and variance $(1-\kappa)\sigmeas^2$ and is
independent of $x_{i,1}$. Selection
acts on $x_{i,1}$ alone. For $Z\sim\mathcal N(0,1)$,
$\mathrm{E}(Z\mid Z<a)=-\lambda(a)$ and
$\mathrm{Var}(Z\mid Z<a)=1-a\lambda(a)-\lambda(a)^2$, so
\begin{align}
\mathrm{E}(\epsilon_{i,1}\mid S_i)
&=-\frac{\sigmeas^2}{\sigtot}\,\lambda(c^{\star}),
\label{eq:s-mean}\\
\mathrm{Var}(\epsilon_{i,1}\mid S_i)
&=\sigmeas^2-\frac{\sigmeas^4}{\sigtot^2}\,
\lambda(c^{\star})\{c^{\star}+\lambda(c^{\star})\}.
\label{eq:s-var}
\end{align}
The factor $\lambda(a)\{a+\lambda(a)\}=1-\mathrm{Var}(Z\mid Z<a)$ lies in
$(0,1)$, decreases in $a$ and tends to one as $a\to-\infty$.

Given $M=m\geq2$ retested individuals, their differences are independent draws
from the conditional distribution of $\Delta_i$ given $S_i$. Their sample mean and sample
variance can be defined using denominator $M-1$, by
\[
\bar\Delta_{\mathcal R}=\frac{1}{M}\sum_{i\in\mathcal R}\Delta_i,
\qquad
s^2_{\Delta,\mathcal R}=\frac{1}{M-1}
\sum_{i\in\mathcal R}(\Delta_i-\bar\Delta_{\mathcal R})^2.
\]
The naive estimator
$\widehat\sigma^2_{\mathrm{meas,naive}}=s^2_{\Delta,\mathcal R}/2$ satisfies
\begin{equation}
\mathrm{E}\bigl(\widehat\sigma^2_{\mathrm{meas,naive}}\mid M=m\bigr)
=\sigmeas^2-\frac{\sigmeas^4}{2\sigtot^2}\,
\lambda(c^{\star})\{c^{\star}+\lambda(c^{\star})\}.
\label{eq:s-naive}
\end{equation}
The bias is negative for every $c$ and smaller in magnitude than
$\sigmeas^4/(2\sigtot^2)$.

For the mean-square estimator, substituting \eqref{eq:s-mean} and
\eqref{eq:s-var} into
\[
\frac{1}{2}\mathrm{E}(\Delta_i^2\mid S_i)
=\frac{1}{2}\mathrm{Var}(\Delta_i\mid S_i)
+\frac{1}{2}\{\mathrm{E}(\Delta_i\mid S_i)\}^2
\]
cancels the terms in $\lambda(c^{\star})^2$, giving, for $m\geq1$,
\begin{equation}
\mathrm{E}\Bigl(\frac{1}{2M}\sum_{i\in\mathcal R}\Delta_i^2\Bigm| M=m\Bigr)
=\sigmeas^2-\frac{\sigmeas^4}{2\sigtot^2}\,c^{\star}\lambda(c^{\star}).
\label{eq:s-meansq}
\end{equation}
This exceeds $\sigmeas^2$ whenever $c^{\star}<0$, equals it when
$c^{\star}=0$, and is smaller when $c^{\star}>0$. Since $\lambda(a)\sim-a$ as
$a\to-\infty$, its conditional bias is asymptotic to
$\{\sigmeas^4/(2\sigtot^2)\}(c^{\star})^2$.

\section{Estimating equation for the correlation}
\label{s:rho}

Fix $\mupop$ and $\sigtot$ at the sample mean and standard deviation of the
first measurements and set $z_{i,j}=(x_{i,j}-\mupop)/\sigtot$. Under the model
of Section~\ref{s:bias}, $(z_{i,1},z_{i,2})$ is standard bivariate normal with
correlation $\rho$. Up to terms
free of $\rho$, the log-likelihood of the retested pairs is
\[
\ell_{\mathcal R}(\rho)=-\frac{M}{2}\log(1-\rho^2)
-\frac{1}{2(1-\rho^2)}\sum_{i\in\mathcal R}
\bigl(z_{i,1}^2-2\rho z_{i,1}z_{i,2}+z_{i,2}^2\bigr).
\]
Write $\hat r=M^{-1}\sum_{i\in\mathcal R}z_{i,1}z_{i,2}$ and
$\hat q=M^{-1}\sum_{i\in\mathcal R}(z_{i,1}^2+z_{i,2}^2)$. Setting
$d\ell_{\mathcal R}/d\rho$ to zero and multiplying by $-(1-\rho^2)^2/M$, which
is nonzero on $(-1,1)$, gives the cubic
\begin{equation}
g(\rho)=\rho^3-\hat r\rho^2+(\hat q-1)\rho-\hat r=0.
\label{eq:s-cubic}
\end{equation}

Since $g(1)=M^{-1}\sum_{i\in\mathcal R}(z_{i,1}-z_{i,2})^2>0$ and
$g(-1)=-M^{-1}\sum_{i\in\mathcal R}(z_{i,1}+z_{i,2})^2<0$ unless all
retested pairs satisfy $z_{i,1}=z_{i,2}$ or all satisfy $z_{i,1}=-z_{i,2}$,
a root exists in $(-1,1)$. The derivative
$g'(\rho)=3\rho^2-2\hat r\rho+\hat q-1$ has minimum $\hat q-1-\hat r^2/3$, so
$g$ is strictly increasing and the root is unique when
$\hat q>1+\hat r^2/3$. Otherwise the estimate is the root that maximizes
$\ell_{\mathcal R}$.

The cutoff is not present in \eqref{eq:s-cubic} and marginal density of
$z_{i,1}$ does not involve $\rho$, so $\ell_{\mathcal R}$ equals up to a constant to the conditional log-likelihood of $z_{i,2}$ given $z_{i,1}$.
Its score has mean zero at the true $\rho$ given any $z_{i,1}$, so
\eqref{eq:s-cubic} is an unbiased estimating equation for every threshold.

\section{Ignorability of the retesting mechanism}
\label{s:ignorability}

Let $R_i$ and $h_2$ be as defined in the main text, and write
$\pi_i=h_2(x_{i,1};\psi)$ and $g(\cdot)=\gmeas(\cdot\mid\thetameas)$. By
assumption $R_i$ is independent of $(T_i,x_{i,2})$ given $x_{i,1}$.

An individual with $R_i=1$ contributes
$g(x_{i,1}-T_i)\,g(x_{i,2}-T_i)\,\pi_i$. For $R_i=0$ the unrecorded $x_{i,2}$
is integrated out. Since $\pi_i$ does not involve $x_{i,2}$ and
$\int g(x-T_i)\,dx=1$, the contribution is $g(x_{i,1}-T_i)(1-\pi_i)$.

Take the prior
$p(\thetapop,\thetameas,\psi)=p(\thetapop,\thetameas)\,p(\psi)$. With
$n_i=1+R_i$, the joint posterior is proportional to $AB$, where
\begin{align*}
A&=p(\thetapop,\thetameas)\prod_{i=1}^{N}\fpop(T_i\mid\thetapop)
\prod_{j=1}^{n_i}g(x_{i,j}-T_i),\\
B&=p(\psi)\prod_{i=1}^{N}\pi_i^{R_i}(1-\pi_i)^{1-R_i}.
\end{align*}
The factor $B$ involves none of $\mathbf T$, $\thetapop$ and $\thetameas$.
Provided $0<\int B\,d\psi<\infty$, integrating out $\psi$ leaves
\[
p(\mathbf T,\thetapop,\thetameas\mid\mathbf x_{\mathrm{obs}},\mathbf R)
\propto A,
\]
which is the posterior used in the main text. The factor $B$ also involves
neither $\fpop$ nor $\gmeas$, so it is the same for every specification that
shares the retesting model and the prior on $\psi$.

\section{Supplementary tables and figures}

\begin{table}[!ht]
\centering
\caption{Simulation settings. Values in g/dL.}
\label{tab:sim-settings}
\begin{tabular}{lp{0.8\textwidth}}
\toprule
Item & Setting \\
\midrule
Model \texttt{a} & $T\sim\mathcal N(15,1)$; $\epsilon\sim\mathcal N(0,0.6^2)$ \\
Model \texttt{b} & $T\sim\mathcal N(15,1)$; $\epsilon\sim t_3(0,0.3464)$ \\
Model \texttt{c} & $T\sim\mathcal N(15,1)$;
$\epsilon\sim0.9\,\mathcal N(0,0.2828^2)+0.1\,\mathcal N(0,1.6971^2)$ \\
Model \texttt{d} & $T\sim\mathrm{SN}(13.7243,1.6210,6)$;
$\epsilon\sim t_5(0,0.4648)$ \\
Priors & Placed on the population mean $\mupop$, population SD $\sigpop$ and
error SD $\sigmeas$.
$\mupop\sim\mathcal N(15,2^2)$; $\sigpop\sim\mathcal N(0,1.5^2)$ and
$\sigmeas\sim\mathcal N(0,1)$ on $[0.02,5]$;
$\log(\nu-2)\sim\mathcal N(\log4,1)$ with $\nu\in[2.1,50]$;
$2w\sim\mathrm{Beta}(2,8)$; $\log(s_2/s_1-1)\sim\mathcal N(\log2,0.8^2)$ with
$s_2/s_1\in[1.05,21]$; $\alpha\sim\mathcal N(0,3^2)$ on $[-10,10]$ \\
Sampling & CmdStan 2.38.0; four chains of 750 warm-up and 1,000 retained
iterations; target acceptance probability 0.97; maximum tree depth 11. Latent
values sampled for models \texttt{a}, \texttt{b} and \texttt{d} and integrated
out analytically for model \texttt{c} \\
Scoring & 400 posterior draws per fit; 200-node Gauss--Hermite quadrature for
models \texttt{b} and \texttt{d}, checked against 300 nodes \\
\bottomrule
\end{tabular}
\end{table}

\begin{table}[!ht]
\centering
\caption{Root mean squared error of the posterior mean and coverage of the 90\%
credible interval, by retest probability $\psi$, over 25 datasets per cell.
The Monte Carlo SE of a coverage of 0.90 is 0.06.}
\label{tab:psi-precision}
\begin{tabular}{llcccc@{\qquad}cccc}
\toprule
& & \multicolumn{4}{c}{Root mean squared error}
& \multicolumn{4}{c}{Coverage}\\
\cmidrule(lr){3-6}\cmidrule(lr){7-10}
Parameter & Model & $0.1$ & $0.5$ & $0.8$ & $1.0$
                  & $0.1$ & $0.5$ & $0.8$ & $1.0$\\
\midrule
\multirow{4}{*}{$\sigmeas$}
 & \texttt{a} & 0.074 & 0.028 & 0.026 & 0.023 & 0.92 & 0.96 & 0.92 & 0.92\\
 & \texttt{b} & 0.109 & 0.052 & 0.055 & 0.058 & 0.80 & 0.96 & 0.88 & 0.88\\
 & \texttt{c} & 0.061 & 0.045 & 0.039 & 0.035 & 0.92 & 0.96 & 0.96 & 1.00\\
 & \texttt{d} & 0.048 & 0.037 & 0.043 & 0.030 & 1.00 & 0.88 & 0.88 & 0.92\\
\addlinespace
\multirow{2}{*}{$\log(\nu-2)$}
 & \texttt{b} & 0.937 & 0.450 & 0.490 & 0.369 & 0.80 & 0.92 & 0.84 & 0.92\\
 & \texttt{d} & 0.601 & 0.444 & 0.482 & 0.534 & 0.92 & 0.96 & 0.92 & 0.84\\
\bottomrule
\end{tabular}
\end{table}

\begin{table}[!ht]
\centering
\caption{Priors for the case-study models. Gamma is parameterized by shape and
rate.}
\label{tab:priors-real}
\begin{tabular}{lp{0.78\textwidth}}
\toprule
Model & Priors \\
\midrule
\texttt{a} & $\mupop\sim\mathcal N(15,2^2)$; $\sigpop,s\sim\mathcal N(0,2^2)$ on
$[0.2,20]$ \\
\texttt{b} & As \texttt{a}; $\nu\sim\mathrm{Gamma}(2,0.1)$ on $[2.01,30]$ \\
\texttt{c} & $\mupop,\sigpop$ as \texttt{a}; $s_1\sim\mathcal N(0,2^2)$ and
$s_2\sim\mathcal N(2,2^2)$ with $0.2\leq s_1\leq s_2\leq2$;
$w\sim\mathrm{Beta}(2,2)$ \\
\texttt{d} & $\xi\sim\mathcal N(15,2^2)$; $\omega\sim\mathcal N(1,2^2)$ on
$[0.2,20]$; $\alpha\sim\mathcal N(0,2^2)$ on $[-5,5]$;
$s\sim\mathcal N(1,2^2)$ on $[0.2,2]$; $\nu$ as \texttt{b} \\
\bottomrule
\end{tabular}
\end{table}

\begin{table}[!ht]
\centering
\caption{Held-out joint ELPD by fold in the case-study comparison. The largest
value in each column is in bold.}
\label{tab:cv-folds}
\begin{tabular}{lrrrrrr}
\toprule
Model & Fold 1 & Fold 2 & Fold 3 & Fold 4 & Fold 5 & Total \\
\midrule
\texttt{a} & -7228.20 & -7166.04 & -7110.03 & -7250.41 & -7168.49 & -35923.17 \\
\texttt{b} & -7192.87 & -7142.56 & -7066.06 & -7079.18 & -7135.55 & -35616.22 \\
\texttt{c} & -7196.86 & -7142.49 & -7068.96 & -7114.04 & -7137.46 & -35659.81 \\
\texttt{d} & \textbf{-7180.25} & \textbf{-7133.72} & \textbf{-7051.23} &
\textbf{-7073.48} & \textbf{-7119.19} & \textbf{-35557.86} \\
\bottomrule
\end{tabular}
\end{table}

\begin{table}[!ht]
\centering
\caption{Specification with the largest held-out ELPD, by generating
specification. Counts are over 50 simulated datasets per row, with the
generating specification in bold.}
\label{tab:selection-matrix}
\begin{tabular}{lrrrr}
\toprule
Generating model & \texttt{a} & \texttt{b} & \texttt{c} & \texttt{d}\\
\midrule
\texttt{a} (normal, normal)   & \textbf{39} & 4 & 0 & 7\\
\texttt{b} (normal, $t$)      & 0 & \textbf{23} & 6 & 21\\
\texttt{c} (normal, mixture)  & 0 & 1 & \textbf{49} & 0\\
\texttt{d} (skew-normal, $t$) & 0 & 0 & 0 & \textbf{50}\\
\bottomrule
\end{tabular}
\end{table}

\begin{table}[!ht]
\centering
\caption{Specification selected by the one-standard-error rule, by generating
specification. Counts are over 50 simulated datasets per row, with the
generating specification in bold.}
\label{tab:selection-one-se}
\begin{tabular}{lrrrr}
\toprule
Generating model & \texttt{a} & \texttt{b} & \texttt{c} & \texttt{d}\\
\midrule
\texttt{a} (normal, normal)   & \textbf{50} & 0 & 0 & 0\\
\texttt{b} (normal, $t$)      & 2 & \textbf{42} & 4 & 2\\
\texttt{c} (normal, mixture)  & 0 & 8 & \textbf{42} & 0\\
\texttt{d} (skew-normal, $t$) & 0 & 0 & 0 & \textbf{50}\\
\bottomrule
\end{tabular}
\end{table}

\begin{table}[!ht]
\centering
\caption{Bias of the posterior mean or median by retest probability $\psi$,
over 25 datasets per cell, with Monte Carlo SE in parentheses. True
$\sigmeas=0.6$ g/dL. Errors and coverage are in Table~\ref{tab:psi-precision}.}
\label{tab:psi-recovery}
\begin{tabular}{llcccc}
\toprule
& & \multicolumn{4}{c}{Retest probability $\psi$} \\
\cmidrule(lr){3-6}
Parameter, generating model & Summary & $0.1$ & $0.5$ & $0.8$ & $1.0$ \\
\midrule
$\sigmeas$, \texttt{a} & mean & $+0.006\;(0.015)$ & $+0.005\;(0.006)$ & $-0.002\;(0.005)$ & $+0.002\;(0.005)$ \\
$\sigmeas$, \texttt{b} & mean & $+0.016\;(0.022)$ & $-0.008\;(0.010)$ & $+0.002\;(0.011)$ & $+0.004\;(0.012)$ \\
$\sigmeas$, \texttt{c} & mean & $+0.017\;(0.012)$ & $-0.009\;(0.009)$ & $-0.008\;(0.008)$ & $-0.018\;(0.006)$ \\
$\sigmeas$, \texttt{d} & mean & $-0.017\;(0.009)$ & $+0.002\;(0.008)$ & $-0.000\;(0.009)$ & $+0.001\;(0.006)$ \\
\addlinespace
$\nu$, \texttt{b} $(\nu=3)$ & mean & $+2.16\;(0.51)$ & $+0.55\;(0.12)$ & $+0.58\;(0.13)$ & $+0.29\;(0.09)$ \\
 & median & $+1.49\;(0.38)$ & $+0.41\;(0.10)$ & $+0.45\;(0.11)$ & $+0.22\;(0.09)$ \\
\addlinespace
$\nu$, \texttt{d} $(\nu=5)$ & mean & $+1.87\;(0.56)$ & $+1.07\;(0.44)$ & $+1.50\;(0.45)$ & $+1.02\;(0.48)$ \\
 & median & $+0.77\;(0.41)$ & $+0.46\;(0.34)$ & $+0.84\;(0.33)$ & $+0.52\;(0.38)$ \\
\bottomrule
\end{tabular}
\end{table}

\begin{table}[!ht]
\centering
\caption{Difference in held-out ELPD between model \texttt{d} and each
alternative over 20,000 donors, with paired SEs. Positive values favor model
\texttt{d}.}
\label{tab:cv_diff}
\begin{tabular}{l r r}
\toprule
Comparison & $\mathrm{ELPD}_{d}-\mathrm{ELPD}_{\mathrm{other}}$ & Paired SE \\
\midrule
\texttt{d} versus \texttt{b} & 58.36 & 11.83 \\
\texttt{d} versus \texttt{c} & 101.95 & 30.69 \\
\texttt{d} versus \texttt{a} & 365.31 & 108.10 \\
\bottomrule
\end{tabular}
\end{table}

\begin{figure}[!ht]
\centering
\includegraphics[width=0.7\linewidth,trim=83 115 83 45,clip]{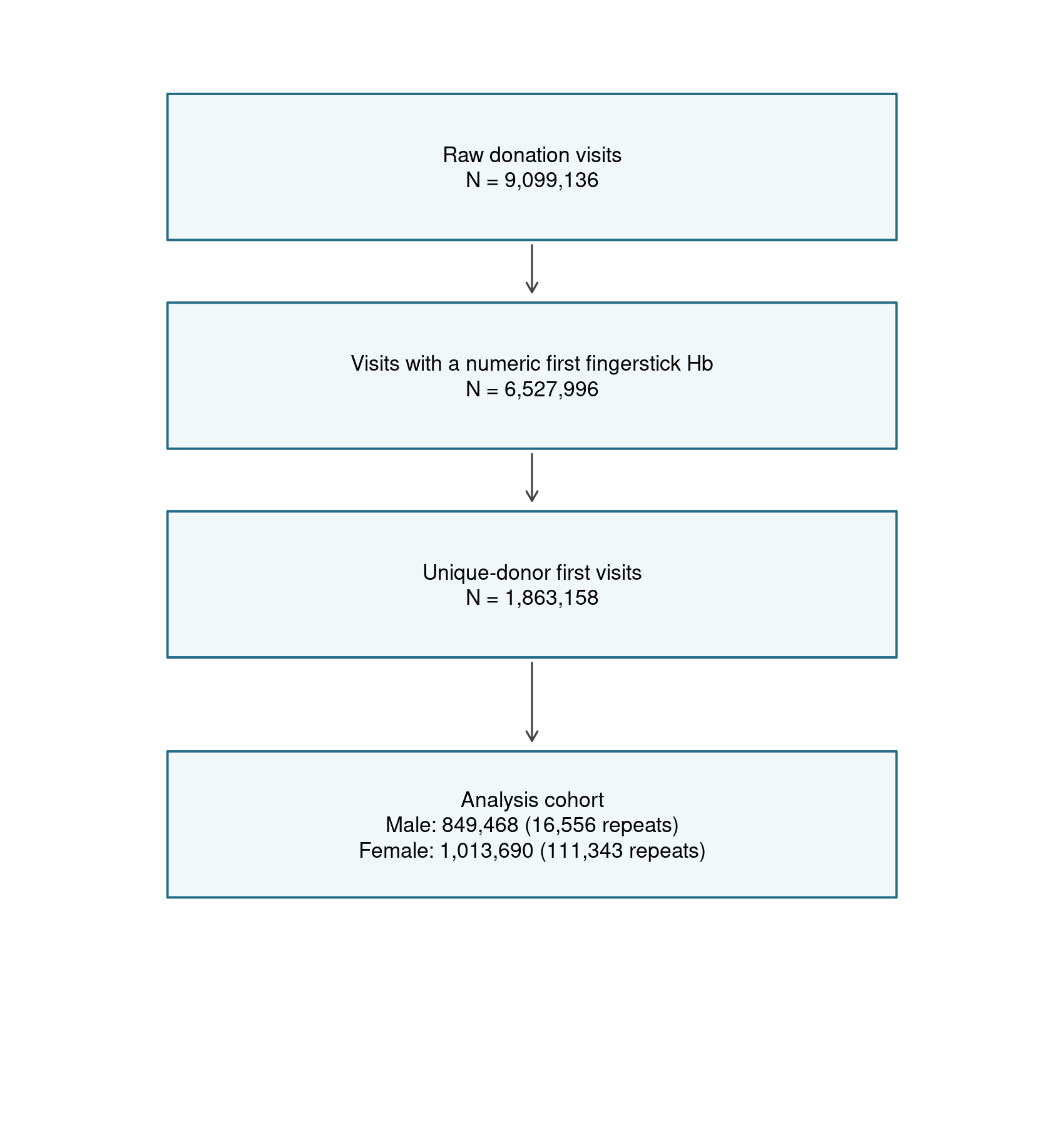}
\caption{Selection of the unique-donor first-visit analysis cohort.}
\label{fig:selection-flow}
\end{figure}

\begin{figure}[!ht]
\centering
\includegraphics[width=0.9\linewidth]{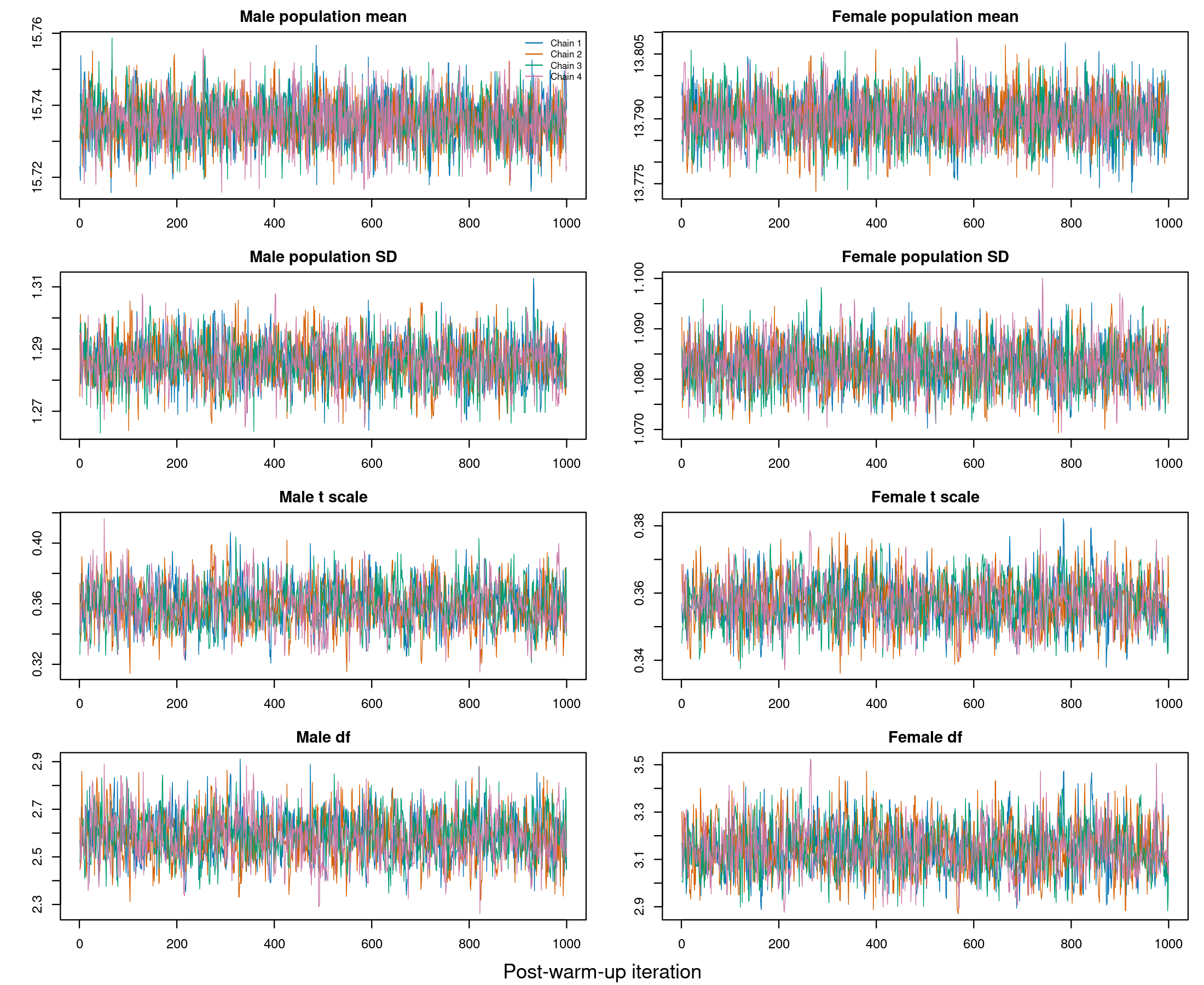}
\caption{Post-warm-up trace plots for the final model \texttt{b} fit. Colors
distinguish the four chains.}
\label{fig:trace}
\end{figure}

\end{appendices}
\label{lastpage}
\end{document}